\documentclass[conference]{IEEEtran}

\ifCLASSINFOpdf
  \usepackage[pdftex]{graphicx}
\else
\fi

\usepackage{siunitx}
\usepackage{cite}
\usepackage[hidelinks]{hyperref}
\usepackage{amsfonts}
\usepackage[inkscapeformat=png]{svg}
\usepackage{diagbox}
\usepackage{colortbl}
\usepackage{multirow}
\usepackage{tabularx}
\usepackage{bm}
\usepackage{textcomp}
\usepackage{multicol}
\usepackage{tikz}
\newcommand\copyrighttext{%
  \footnotesize \textcopyright 2025 IEEE. Personal use of this material is permitted.
  Permission from IEEE must be obtained for all other uses, in any current or future
  media, including reprinting/republishing this material for advertising or promotional
  purposes, creating new collective works, for resale or redistribution to servers or
  lists, or reuse of any copyrighted component of this work in other works.
  DOI: \href{https://ieeexplore.ieee.org/document/11037177}{10.1109/EuCNC/6GSummit63408.2025.11037177}}
\newcommand\copyrightnotice{%
\begin{tikzpicture}[remember picture,overlay]
\node[anchor=south,yshift=10pt] at (current page.south) {\fbox{\parbox{\dimexpr\textwidth-\fboxsep-\fboxrule\relax}{\copyrighttext}}};
\end{tikzpicture}%
}

\begin{document}
%
%\title{CSI-Based Cross-Technology Interference Detection and Classification for Wi-Fi 6 OFDMA Scheduling}
\title{Wi-Fi 6 Cross-Technology Interference Detection and Mitigation by OFDMA: an Experimental Study}
% \title{Cross-Technology Interference Detection and Mitigation by Wi-Fi 6 OFDMA: Experimental Study}
%\title{Experimental study on Cross Technology Interference Detection and Mitigation using OFDMA}

% Cross-Technology Interference Detection on Wi-Fi 6 \\ IoT Hardware Enabling OFDMA-based Mitigation
% CSI-Based Cross-Technology Interference Detection and Classification for Wi-Fi 6 OFDMA Scheduling
% Cross-Technology Interference Detection on Wi-Fi 6 IoT Hardware for OFDMA-based Mitigation 

% author names and affiliations
% use a multiple column layout for up to three different
\author{\IEEEauthorblockN{Thijs Havinga, Xianjun Jiao, Wei Liu, Baiheng Chen, Adnan Shahid and Ingrid Moerman}
\IEEEauthorblockA{IDLab, \textit{Department of Information Technology},
Ghent University - imec\\
Ghent, Belgium\\
\{Thijs.Havinga, Xianjun.Jiao, Wei.Liu, Colvin.Chen, Adnan.Shahid, Ingrid.Moerman\}@UGent.be}
}

% conference papers do not typically use \thanks and this command
% is locked out in conference mode. If really needed, such as for
% the acknowledgment of grants, issue a \IEEEoverridecommandlockouts
% after \documentclass

% make the title area
\maketitle

% As a general rule, do not put math, special symbols or citations
% in the abstract
\begin{abstract}
Cross-Technology Interference (CTI) poses challenges for the performance and robustness of wireless networks. There are opportunities for better cooperation if the spectral occupation and technology of the interference can be detected. Namely, this information can help the Orthogonal Frequency Division Multiple Access (OFDMA) scheduler in IEEE 802.11ax (Wi-Fi 6) to efficiently allocate resources to multiple users in the frequency domain. 
This work shows that a single Channel State Information (CSI) snapshot, which is used for packet demodulation in the receiver, is enough to detect and classify the type of CTI on low-cost Wi-Fi 6 hardware. We show the classification accuracy of a small Convolutional Neural Network (CNN) for different Signal-to-Noise Ratio (SNR) and Signal-to-Interference Ratio (SIR) with simulated data, as well as using a wired and over-the-air test with a professional wireless connectivity tester, while running the inference on the low-cost device. Furthermore, we use \textit{openwifi}, a full-stack Wi-Fi transceiver running on software-defined radio (SDR) available in the w-iLab.t testbed, as Access Point (AP) to implement a CTI-aware multi-user OFDMA scheduler when the clients send CTI detection feedback to the AP. We show experimentally that it can fully mitigate the 35\% throughput loss caused by CTI when the AP applies the appropriate scheduling.%, whereas using a naive scheduler it leads to 35\% throughput reduction.
\end{abstract}
\IEEEpeerreviewmaketitle
\copyrightnotice

\vspace*{-1em}  % Tweak this value as needed 
\section{Introduction and related work}
Partly thanks to the rise of the Internet-of-Things (IoT), there is an ongoing increase of wireless communication devices operating in unlicensed bands. Therefore, Cross-Technology Interference (CTI) becomes an increasingly important problem.
%Furthermore, interference is not limited to devices within a single vehicle: multiple in-vehicular networks can also cause CTI between different vehicles, as highlighted by \cite{Na16} and \cite{Veisi20}.
Particularly in the 2.4GHz band, a multitude of technologies coexist, such as Wi-Fi (IEEE 802.11), Bluetooth/Bluetooth Low Energy (BLE), and Low-Rate Wireless Personal Area Network (LR-WPAN) devices based on the IEEE 802.15.4 standard like ZigBee, WirelessHART or Thread \cite{Nikoukar18}. 
Although newer Wi-Fi standards can also operate in the 5GHz and 6GHz bands, the 2.4GHz band is still actively used due to its higher coverage range and compatibility with legacy devices. In our previous work \cite{Havinga24}, we demonstrated that even at a high Signal-to-Noise Ratio (SNR) of \SI{23}{dB}, CTI can significantly impact the Packet Error Rate (PER) of Wi-Fi. When employing a high Modulation and Coding Scheme (MCS) such as MCS 7, even relatively low interference power from ZigBee that results in an Signal-to-Interference Ratio (SIR) of approximately \SI{15}{dB}, can lead to 100\% PER. 

%Thread is a rather new mesh networking standard, which is used for home automation with the Matter standard. The latest iPhones and other Apple devices are also equipped with such a radio \cite{Verge}, suggesting that this standard is on the path to becoming widely adopted. 

There are a few methods to manage spectrum access under CTI. Several technologies apply a Clear Channel Assessment (CCA), usually consisting of an energy detection and a signal detection step. Energy detection only works when the interference is significantly strong. When it is performed on different bandwidths with different thresholds, this may lead to asymmetric access opportunities. %The same holds for systems with different radio turnaround times; if one technology determines the channel is idle, but takes a longer time to access the channel, another technology might jump in.
Moreover, energy detection at the transmitter may not be a good indicator of the CTI level perceived by the receiver, causing the hidden or exposed node problem.
Lastly, since signal detection\textemdash usually performed by correlating the incoming signal with a known sequence\textemdash is technology-specific, it does not work for CTI \cite{Nikoukar18}. 

Other CTI detection methods rely on high-level metrics like PER or receiver error codes \cite{Croce18}. However, since \cite{Croce18} relies on the temporal gap between errors, a fixed packet size and interval are assumed. Even if this assumption is met, the spectral location of CTI cannot be determined.
%The work in \cite{Michelson19} uses channel state flags reported by the receiver to detect interference in IEEE 802.11p networks, which works well in case of high interference power, but the detection rate quickly drops below 50\% for an interference power lower than \SI{-70}{dBm}.

% The authors of \cite{Peha24} propose certain coexistence rules to mitigate interference between Cellular V2X communications and Wi-Fi. They propose to let V2X devices send Wi-Fi compliant Clear-to-Send packets and create on-off periods where only one of the technologies is allowed to transmit. While this allows to allocate resources according to the on-off period, it can still have a significant impact on the performance of either technology. Furthermore, it requires the V2X devices to be able to send a Wi-Fi-compliant packet.

%\subsection{Technology recognition}
Technology recognition can be used to improve networks using domain-specific knowledge about the detected technology. 
For example, while ZigBee uses a fixed channel, BLE uses a frequency hopping scheme. To mitigate ZigBee, knowing the spectral location can be used to avoid future interference, while for BLE real-time detection is needed for each transmission to overcome an ongoing interference packet. % Knowing the properties of the coexisting technologies may also overcome the issue of starvation caused by asymmetric CCA mechanisms. 
While technology recognition can be valuable, it typically requires additional hardware, such as a separate software-defined radio acting as sensing engine \cite{Fontaine20}. %, Girmay2023 
Alternatively, communication devices may perform a low-resolution spectrum scan using existing hardware, but it requires an idle period, as demonstrated by \cite{Chwalisz18}. These limitations make it impractical for low-cost deployments, which we will address in this work.

% \subsection{CSI-based spectrum sensing}
Channel State Information (CSI) is derived from the Fast Fourier Transform (FFT) of the training sequence in a Wi-Fi packet, which is required to demodulate the packet. Hence, it can also act as a spectrum sensing module without extra hardware cost.
%It reveals the propagation conditions of the environment, from which also other aspects can be inferred, such as human activity recognition \cite{ARMENTAGARCIA2024254}. 
Several Wi-Fi chips expose the CSI to the user, including the low-cost ESP32-C6 \cite{ESP32-C6}, which we use in this work. 
However, there is little existing work on using CSI for interference detection, which we discuss hereafter.

The work in \cite{Zheng2017DetectingRF} detects interference of Bluetooth, ZigBee and microwave ovens using the CSI of Wi-Fi devices with 52 subcarriers. In their subsequent work \cite{Zheng2018}, they also identify which subcarriers are interfered by estimating the distortion peak, which is assumed to be at the center of a fixed bandwidth. In static environments using 500 CSI snapshots, ZigBee and Bluetooth are correctly detected in more than 90\% of the time with almost 100\% true negative rate, while in dynamic environments it drops to about 86\%. In both cases, the interferer was at \SI{1}{m} distance from the Wi-Fi receiver, while at \SI{4}{m} the ZigBee detection rate dropped to 25\%.

The authors of \cite{Yang18} use different machine learning methods to classify interference of Bluetooth, Wi-Fi and microwave ovens using a Wi-Fi chip which provides 30 CSI values per snapshot. By creating heavy Wi-Fi traffic such that during a sample period of \SI{1}{s} 400 CSI snapshots are recorded, the classification accuracy of the best method reaches around 90\% accuracy, but the SNR and SIR are not specified.\\

% AiFi \cite{Chen23} uses pilot information and CSI to improve the channel equalization and decoding process in the presence of interference using neural networks. Rather than technology classification, their output is the estimated interference signal that is used to clean the signal in a later stage.
% Another benefit of using CSI for interference detection is that it contains information about the interference strength as well.
% The authors of \cite{Zhang21} have shown that CSI can also be used to approximate the location of an interferer in a classroom with an average accuracy of about \SI{1.5}{m} using 100 snapshots. Data is collected with known interference location in a stable environment, after which the methods dynamic time warping and gradient boosting regression tree are used to localize the interferer.

Orthogonal Frequency Division Multiple Access (OFDMA) is introduced to Wi-Fi in the IEEE 802.11ax standard to serve multiple users at the same time using different subcarrier sets, called Resource Units (RUs). This reduces the time spent for contention to access the channel and limits the overhead of the preamble, thereby lowering latency and increasing throughput. The standard defines the necessary guidelines to ensure interoperability, but the scheduling of the RU width (number of subcarriers), RU index (spectral location) and MCS to the different users, etc., is left to vendors to implement. Several works consider OFDMA scheduling while taking into account the channel variation across the occupied spectrum, mainly due to small-scale fading. The authors of \cite{Wang20} use a per-subcarrier channel gain to maximize the overall data rate by allocating RUs and MCSs appropriately. The work in \cite{Tutelian21} includes channel fading from CSI feedback into the scheduler for allocating RUs and MCSs. While this is a valid approach in interference-free channels, CTI usually constitutes as a boost in the magnitude of the CSI, while the packet reception performance is actually worse. In our previous work \cite{Havinga24}, we have shown the concept of puncturing an RU\textemdash meaning no data is sent on subcarriers belonging to a certain RU\textemdash when affected by CTI in a single-user (SU) scenario. There it is assumed that the interference is known to the AP. In this work, we realize the CTI detection feedback and we create a CTI-aware multi-user (MU) OFDMA scheduler as explained next.

Since Wi-Fi 6, packets are modulated on a maximum of 242 instead of 52 active OFDM subcarriers per 20MHz bandwidth for 802.11a/g, or 56 for 802.11n/ac. The CSI in Wi-Fi 6 is the FFT of the High Efficiency Long Training Field (HE-LTF) in the preamble, after compensating for the known sequence. The highest frequency resolution is achieved with a \SI{12.8}{\micro \second} long HE-LTF, which consists of 242 active subcarriers, which is more than four times the amount of earlier standards. This allows us to classify CTI with only one CSI snapshot obtained from regular Wi-Fi traffic on low-cost hardware, eliminating the need to transmit multiple CSI snapshots to a high-end AP for processing. % This is beneficial, since a high CSI sampling rate is often not feasible, because Wi-Fi packets last anywhere from \SI{24}{\micro s} (preamble and 1 OFDM symbol) to \SI{5.484}{ms}, which is the maximum allowed by the 802.11ax standard. Furthermore, there are mandatory inter-frame spaces and the need to apply listen-before-talk. 

Since commercial Wi-Fi 6 APs lack control over low-level OFDMA features, we use openwifi \cite{jiao2020openwifi} to implement a CTI-aware scheduler. We show that with spectral knowledge of the CTI, for MU packets, an affected RU can be assigned to a user that experiences less CTI due to its physical location, and this improves the network's throughput.

To the best of our knowledge, this work is the first to achieve the following combination of attractive properties:
\begin{enumerate}
    \item Interference detection including spectral location and technology classification; 
    \begin{itemize}
        \item Requiring only a single CSI snapshot obtained from an existing FFT module in the Wi-Fi chipset;
        \item Executable on very low-cost ($\approx$ \$8) hardware.
    \end{itemize}
    \item A CTI-aware OFDMA scheduler using CSI-based technology classification demonstrating to mitigate 35\% loss in throughput due to CTI.
\end{enumerate}

% Although this work focuses on classifying CTI from common in-vehicle devices using LR-WPAN or BLE on the 2.4GHz band, it can be extended to other technologies as well.\\
% Furthermore, we show what the influence of different SNR and SIR is on the classification accuracy and give useful insights obtained from real-life experiments.

% The paper is structured as follows. First, we present the system model for CTI classification using CSI. Then we describe the methodology to implement the model, after which we evaluate the performance of the classification model. Next, we discuss the implementation of the CTI-aware OFDMA scheduler on an SDR, and assess its performance using a real-life experiment. Lastly, we conclude the paper and give some suggestions for future work.

\section{System Model}
\label{sec:theory}
Given a known transmitted OFDM signal in the frequency domain $\bm{x}$ (namely the HE-LTF in 802.11ax), a transmitted interference signal $\bm{i}$ and noise vector $\bm{n}$, the received OFDM symbol $\bm{y}$ in a frequency selective fading environment is then formulated as:
\begin{equation}
    \bm{y} = \bm{h_w} \odot \bm{x} + \bm{h_i} \odot \bm{i} + \bm{n},
\end{equation}
where $\bm{h_w}$ and $\bm{h_i}$ are the channel responses of the Wi-Fi and interference signal, respectively, and $\odot$ represents element-wise multiplication. All signals are vectors of complex values with dimension $M$ as the number of active subcarriers of the Wi-Fi OFDM symbol.
The observed CSI $\bm{\hat{h}}$ by a Wi-Fi receiver is then formulated as: 
\begin{equation}
\label{eq:csi}
    \bm{\hat{h}} = \bm{y} \oslash \bm{x} = (\bm{h_w} \odot \bm{x} + \bm{h_i} \odot \bm{i} + \bm{n}) \oslash \bm{x},
\end{equation}
where $\oslash$ represents element-wise division. From the characteristics of $\bm{i}$, such as its presence on different subcarriers, as well as its magnitude and phase change due to different modulation types, the spectral location and technology can be derived. Given a newly observed CSI $\bm{\hat{h}}$, the task is thus to estimate $\bm{i}$ and extract features from $\bm{i}$ to classify the type of CTI, if any. This is a typical task for a Convolutional Neural Network (CNN). By training it on different $\bm{\hat{h}}$ with and without interference, the model learns each of the individual components of the received signal in order to determine $\bm{i}$ from a new CSI $\bm{\hat{h}}$. The benefit of using machine learning over purely rule-based methods is that it does not require threshold tuning and it can easily be retrained with real-life data.  

\begin{figure}
    \centering
    \includegraphics[width=\linewidth]{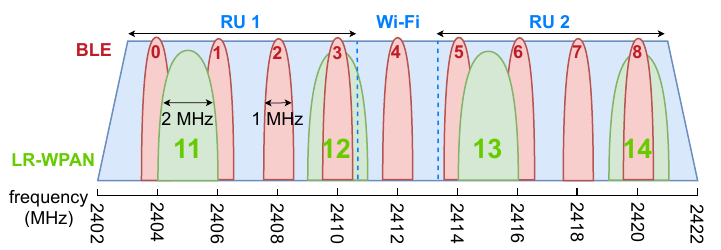}
    \caption{Spectrum used by Wi-Fi channel 1 (dotted lines show borders between two 106-tone RUs), LR-WPAN channels 11-14 and BLE channels 0-8.}
    \label{fig:BLE_LRWPAN}
\end{figure}

%In principle, any narrowband interferer in overlapping spectrum can be detected using CSI, and hence mitigated with OFDMA scheduling. 
A typical scenario where this theory can be applied is for a Wi-Fi channel in the \SI{2.4}{GHz} ISM band, which overlaps with multiple LR-WPAN and BLE channels, %(using the default \SI{1}{MHz} bandwidth), 
as shown in Fig. \ref{fig:BLE_LRWPAN}. 
Before Wi-Fi 6, the entire bandwidth (indicated by blue colour) is either occupied or free; with OFDMA enabled, Wi-Fi 6 allows dividing subcarriers into several RUs, the boundaries of the RUs comprised of 106 subcarriers (106-tone RUs) are highlighted by dashed blue lines.

\begin{figure}
    \centering
    \includegraphics[width=\linewidth]{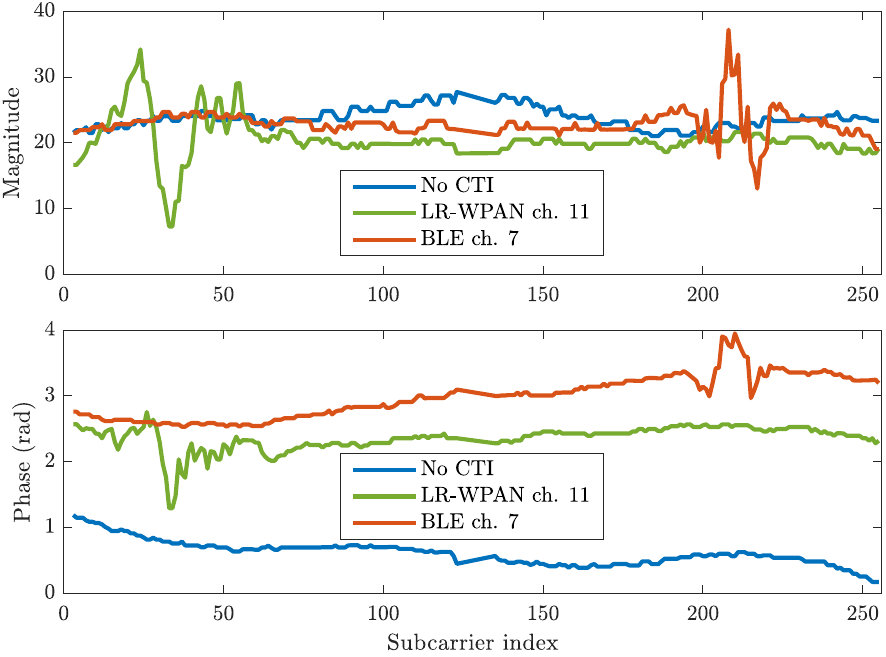}
    \caption{Magnitude and phase of CSI snapshots captured at high SNR by the ESP32-C6 when not interfererd (blue), interfered with LR-WPAN channel 11 (green), or BLE channel 7 (red) with an SIR of \SI{1}{dB}.}
    \label{fig:CSI}
\end{figure}

Fig. \ref{fig:CSI} shows the magnitude and phase of three typical CSI snapshots captured by ESP32-C6 on Wi-Fi channel 1, either with no interference, or with CTI from LR-WPAN or BLE. % The blue snapshot contains no interference, the green one is captured when a LR-WPAN signal is present on channel 11 and the red one is obtained with BLE on channel 7, both with an SIR of \SI{1}{dB}.
In real-life scenarios, packets with different SNR and SIR levels will be received. The SNR represents the power of $\bm{h_w} \odot \bm{x}$ as compared to $\bm{n}$ and thus influences the quality of the CSI. Similarly, the SIR refers to the power ratio of $\bm{h_w} \odot \bm{x}$ with respect to $\bm{h_i} \odot \bm{i}$. The higher the SIR, the weaker the interference signal, the more difficult it is to detect and characterize CTI. 

%In \cite{Havinga24}, we showed that an SIR of more than \SI{20}{dB} can still influence the PER of Wi-Fi traffic with high MCS, while the interference is likely not detected by CCA. Hence, it is important to perform CTI detection and recognition even at high SNR or SIR.

\section{Methodology and proposed CNN}
\begin{figure}[h]
    \centering
\includegraphics[width=\linewidth]{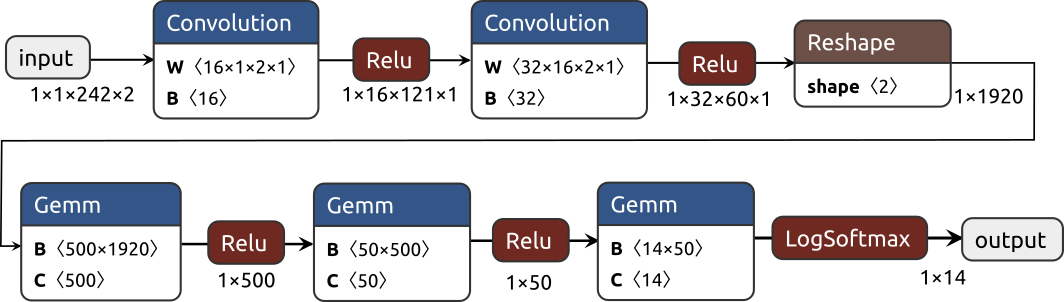}
    \caption{Diagram of the CNN with number of \textbf{W}eigths, \textbf{B}iases and \textbf{C}hannels.}
    \label{fig:CNN}
\end{figure}

%This section presents the methodology, in which we discuss the structure of the CTI classification model, the data generation, the training and deployment of the model.
We create a CNN that has a matrix input $\mathbf{\hat{H}} \in \mathbb{R}^{2 \times 242}$, representing the real and imaginary part of the 242 active subcarriers of the CSI. Note that the CSI is thus not converted to magnitude and phase components to avoid this computational step on-device. The output is 14 classes $\{C_{1},C_{2},\dots, C_{14}\}$, where $C_{1}$ is no interference, $C_2$ to $C_{5}$ are the interference of the LR-WPAN channels and the remaining classes, $C_6$ to $C_{14}$, pertain to the interference of the BLE channels (see Fig. \ref{fig:BLE_LRWPAN}). In order to keep the size of the CNN small, and since there are less features to be extracted compared to e.g. image recognition, it consists of only two convolutional layers. Afterwards, in order to understand the complex relationships between features like the shape of the overall CSI and the distortions due to interference, three fully-interconnecting layers with generalized matrix multiplication (Gemm) are applied. Each layer uses Rectified Linear Unit (ReLU) activation and at the final stage, a LogSoftmax function generates the probability distribution across the classes. See Fig. \ref{fig:CNN} for the full model.

In order to create a general model that is not dependent on specific transmitter, receiver or environmental characteristics, we opt to use artificial data to train the model. The data is generated based on the physical layer specifications and multipath channel models of the relevant standards.
Namely, we prepare a 242-tone HE-LTF field in MATLAB and generate several LR-WPAN and BLE waveforms using the Communications and Bluetooth Toolbox, respectively. % These signals are then sampled at \SI{20}{MHz}, which is according to the Wi-Fi standard. %The LR-WPAN signal has an effective bandwidth of \SI{2}{MHz}, while that of BLE is \SI{1}{MHz}. 
We shift the LR-WPAN and BLE signals in the frequency domain to one of the overlapping channels within Wi-Fi channel 1
%channels 11-14 of LR-WPAN and channels 0-8 of BLE, which overlaps with Wi-Fi channel 1 
as shown in Fig. \ref{fig:BLE_LRWPAN}. 
The HE-LTF and the interfering signals are convoluted with a different realization of the IEEE 802.11ax indoor spatial channel models B or C \cite{TGax}, suitable for small to medium indoor environments. No carrier and sampling frequency offset are added to the HE-LTF symbol, because the receiver will compensate for this already using the legacy preamble training fields.
Afterwards, we scale the IQ samples of the interference according to a certain SIR of the full \SI{20}{MHz} and superpose them with the HE-LTF. Then we add white Gaussian noise according to the SNR, which is based on the signal power without channel response and interference. Following, we perform the 256-point FFT on it, and compensate it with the known HE-LTF sequence to get the CSI at the active subcarriers. 

The output of CSI of the ESP32-C6 are two 8-bit signed integers per subcarrier, one for the real part and one for the imaginary part. The typical range of the magnitude of the CSI is experimentally determined by receiving packets from the highest to lowest decodable RSSI range. The CSI output from the generated data is scaled such that the average magnitude is in this range. For each SIR and SNR value (as determined below), we generate 2000 CSI snapshots with random noise and realization of the channel model, and let a random part of the CTI signals overlap with the HE-LTF.

The CNN is implemented in Python using the PyTorch library. Models are initially trained based on CSI obtained at specific SNR and SIR to determine the range at which sufficient accuracy is achieved. The general model is trained on data with an SNR range of \SIrange{14}{24}{dB} and an SIR range of \SIrange{1}{15}{dB}, varied in \SI{1}{dB} steps. 
%At the starting point, 80\% classification accuracy is achieved at moderate SNR and SIR (i.e.,SNR \SI{14}{dB}, SIR \SI{15}{dB}). It drops sharply at lower SNR, especially when also combined with high SIR (weaker interference). On the contrary, for strong signal with strong interference (eg, SNR \SI{24}{dB}, SIR \SI{1}{dB}), the CSI quality is very good, but packets are often not decoded. Thus, CSI data can be rarely obtained on the Wi-Fi receiver, which leads to low detection accuracy as well. 
The generated data is divided into 80\% training and 20\% validation data. We use the Adam optimizer with a learning rate of 0.001 and a batch size of 256 for 200 epochs with negative log likelihood loss function. Training was performed on an 8-core Intel Xeon CPU E5-2620 v4 at 2.10GHz with 32GB RAM, which took about 10 hours. We convert the PyTorch model to the Open Neural Network Exchange (ONNX) format. It is then pre-processed and optimized without quantization for the ESP32. %using the ESP-DL framework. 
%This utilizes the Apache TVM framework to convert the model to be run by a C++ runtime. 
% The inference is executed in a callback when a CSI snapshot is received. Note that the chip can be set to promiscuous mode in order to get CSI snapshots even when the packet is not addressed to it, which effectively increases the amount of CSI data collected for CTI classification. % One caveat is that the ESP32-C6 seems to report CSI only when the packet was successfully decoded, meaning that a low MCS is needed to overcome strong interference.

The firmware is open source\footnote{Link to the source code for training the model and the firmware: \href{https://github.com/HavingaThijs/CSI-Based_CTI_Detection}{\ttfamily\url{https://github.com/HavingaThijs/CSI-Based\_CTI\_Detection}}.} and consumes 4.77MB of the available 8MB flash memory on the ESP32, of which 0.91MB is occupied by the ESP-IDF framework including Wi-Fi driver.
The model runs with a clock frequency of \SI{160}{MHz} on the ESP32-C6 single-core RISC-V RV32IMAC CPU, which lacks dedicated floating-point operations. The result is a considerable inference time (i.e. \SI{680}{ms}). Techniques such as removing packets at too low SNR, applying integer quantization, or decreasing the model size when only the spectral location needs to be detected, can be used to reduce the inference time.

\section{Performance evaluation of CTI detection}
\begin{figure*}[t]
    \centering
    %\includesvg[width=\textwidth]{images/accuracy.svg}
    \includegraphics[width=\textwidth]{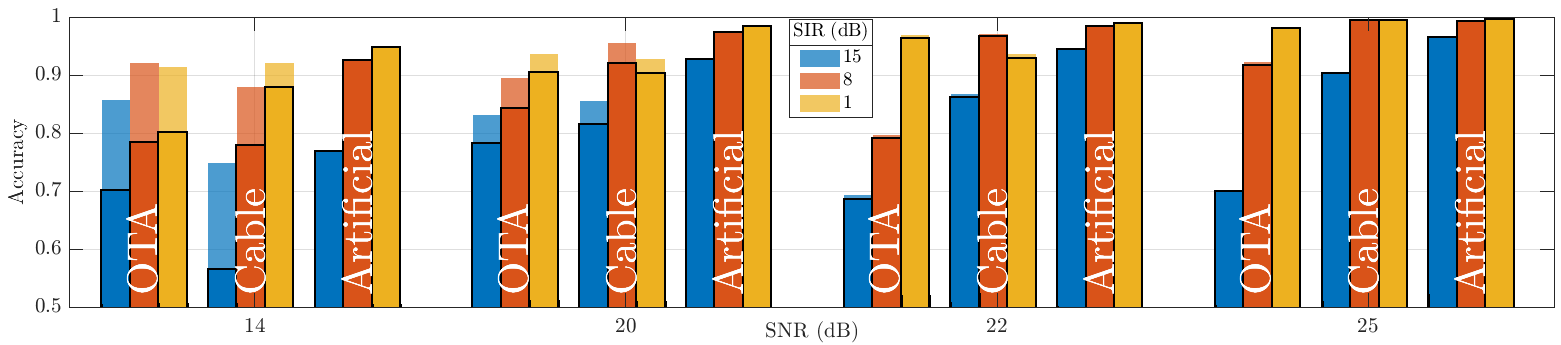}
    \caption{Classification accuracy of the OTA, cable and artificial tests for different SNR and SIR (note that the accuracy starts at 0.5 for better visibility).}
    \label{fig:accuracy}
\end{figure*}

Three different tests are done to evaluate the performance of the classification model, namely an over-the-air (OTA) test, a test using coaxial cable, and a test with artificial data on a host PC. The OTA test includes hardware impairments as well as effects from unseen channel responses. The cable test eliminates the channel response, while the artificial data shows the performance without any hardware impairments on a non-constraint device. % Hence the tests are complementary, they increase the completeness and the confidence of the evaluation process.

%Two major challenges are faced in the OTA test: (i) real LR-WPAN or BLE devices transmits intermittently, it is difficult to determine which packets' CSI contain interference; (ii) many uncontrolled devices can generate signals in unlicensed bands, there is a high risk to have a contaminated result.  
In order ensure 100\% interference ratio, we utilize the R\&S\textsuperscript{\textregistered}CMW270 wireless connectivity tester as an IEEE 802.11ax AP, which simultaneously replays clean IQ samples of the waveforms of LR-WPAN channels 11-14 or BLE channels 0-8 via the general purpose RF output. The AP is set to use channel 1 and create HE-SU packets using MCS 0. The Wi-Fi tester is set to its maximum transmit power of \SI{-3}{dBm} with antennas connected, while the ESP32-C6's antenna port is connected to an RF terminator. In this way, the ESP32-C6 is only able to receive the strong signal from the nearby Wi-Fi tester. 
By moving the ESP32-C6 around in approximately a meter range, we control the SNR relative to the AP's power. Still in some cases a clear unintentional interference was visible in the CSI. These measurements were manually discarded. The SNR is estimated based on RSSI reports during the measurement and the receiver sensitivity of the ESP32-C6. SIR is determined by the power difference between the RF ports for Wi-Fi and the interference signal. For the cable test, the ESP32-C6's antenna port %with U.FL connector 
is connected to the output of a 2-way power combiner, one input is connected by coaxial cable to the tester's RF port for Wi-Fi and one to the RF port for interference. The different RSSI values are obtained by changing the tester's output power. When deploying the model on the ESP32-C6, it could be seen that at lower SNR, the model often judges interference at BLE channel 4, which is exactly in the middle of Wi-Fi channel 1. A possible reason for this is that leakage of the transmitter's or receiver's Local Oscillator (LO) can create distortion in this region. %It may therefore be beneficial to train the model using captured data that incorporates these non-idealities. 
From here on, we use \textit{filtered} accuracy to refer to results without misdetection at BLE channel 4.

We evaluate the performance by running the model on 100 filtered CSI snapshots per class (i.e., no interference, and LR-WPAN or BLE present on one of the overlapping channels with Wi-Fi), at different SNR and SIR on the ESP32-C6. In total, 16,800 filtered snapshots for both the OTA and cable test are collected. The average accuracy is determined by ratio of correctly classified interference source out of all snapshots. Finally, we ran the model on a PC with the artificial test set, using the same SNR and SIR values.

Fig. \ref{fig:accuracy} shows the accuracy of CSI-based technology classification for each combination of SNR and SIR of the OTA, cable and artificial test. For the OTA and cable test we show both the original accuracy (bars with borders) and filtered accuracy (bars without borders). %It can be seen that at SNR above \SI{20}{dB} and an SIR of \SI{8}{dB} and lower, up to 99\% precision can be obtained for both the cable and artificial test, and for the OTA test it is above 90\% for an SNR of \SI{25}{dB}. 
In general, the lower the SNR and the weaker the interference (higher SIR), the lower the accuracy, as expected from Equation \ref{eq:csi}. Two interesting outliers from this general rule can be identified. 
%For the OTA test, the higher the SNR, there is a downward trend in accuracy for an SIR of \SI{15}{dB}. 
The accuracy of the OTA test at SIR \SI{15}{dB} peaked at \SI{20}{dB} SNR; there is a downward trend at higher SNR. This may be because the model has learned that for a smooth CSI (often happens at high SNR), there needs to be a clear distortion visible to classify it as interference. However, in the dynamic environment with unseen channel responses, weak CTI might be easily confused as part of the fading channel. %Indeed, in the OTA test the model seems to be less sensitive to the CTI as can be seen in Table \ref{tab:conftech}, which we discuss later. 
Furthermore, in the cable test at \SI{20}{dB} SNR and above, detecting weaker interference (SIR of \SI{8}{dB}) is better than detecting stronger interference (SIR of \SI{1}{dB}). A reason for this might be that due to the very strong CTI, the synchronization, frequency or sampling offset estimation of the Wi-Fi receiver is more likely to be wrong, which will hurt the quality of the CSI, hence affecting the accuracy.

For the prior work relying on multiple CSI snapshots, \cite{Yang18} does not specify any signal quality or interference power, and \cite{Zheng2018} only specifies distances between the Wi-Fi devices and interferer. Assuming \SI{20}{dBm} Wi-Fi transmit power with the largest distance between Wi-Fi devices of \SI{4}{m} in \cite{Zheng2018}'s experiment, it would lead to an SNR much higher than \SI{25}{dB}. Then they achieve an accuracy higher than 90\% when the ZigBee interferer is located at \SI{1}{m}. Our OTA test achieves similar performance under more difficult conditions (see Fig. \ref{fig:accuracy} at SNR \SI{25}{dB} and SIR \SI{8}{dB}). However, when the ZigBee interferer is placed \SI{4}{m} away from Wi-Fi receiver, their true positive rate already drops to 25\%. This corresponds to roughly \SI{20}{dB} SIR assuming ZigBee's transmit power is \SI{0}{dBm}, whereas in our OTA test, with \SI{25}{dB} SNR and \SI{15}{dB} SIR, the model still maintains 70\% recognition precision.

\begin{table}
\caption{Confusion matrix per technology}
\centering
\begin{tabular}{|ll|l|l|l|}
\hline
\multicolumn{2}{|l|}{\diagbox[width=30mm]{\textbf{Actual}}{\textbf{Predicted}}}            & \textbf{No CTI}                                     & \textbf{LR-WPAN}              & \textbf{BLE}                                        \\ \hline
\multicolumn{1}{|l|}{}                                   & \textbf{OTA}   & \cellcolor[HTML]{C0C0C0}94.75                       & 3.0                           & 2.25                                                \\ \cline{2-5} 
\multicolumn{1}{|l|}{}                                   & \textbf{Cable} & \cellcolor[HTML]{C0C0C0}{\color[HTML]{333333} 91.5} & 3.75                          & 4.75                                                \\ \cline{2-5} 
\multicolumn{1}{|l|}{\multirow{-3}{*}{\textbf{No CTI}}}  & \textbf{Artificial}  & \cellcolor[HTML]{C0C0C0}85.53                       & 5.03                          & 9.44                                                \\ \hline
\multicolumn{1}{|l|}{}                                   & \textbf{OTA}   & 15.17                                               & \cellcolor[HTML]{C0C0C0}81.46 & 3.38                                                \\ \cline{2-5} 
\multicolumn{1}{|l|}{}                                   & \textbf{Cable} & 7.54                                                & \cellcolor[HTML]{C0C0C0}87.5  & 4.91                                                \\ \cline{2-5} 
\multicolumn{1}{|l|}{\multirow{-3}{*}{\textbf{LR-WPAN}}} & \textbf{Artificial}  & 2.27                                                & \cellcolor[HTML]{C0C0C0}95.82 & 1.91                                                \\ \hline
\multicolumn{1}{|l|}{}                                   & \textbf{OTA}   & 9.79                                                & 1.8                           & \cellcolor[HTML]{C0C0C0}88.42                       \\ \cline{2-5} 
\multicolumn{1}{|l|}{}                                   & \textbf{Cable} & 2.42                                                & 3.68                          & \cellcolor[HTML]{C0C0C0}{\color[HTML]{333333} 93.9} \\ \cline{2-5} 
\multicolumn{1}{|l|}{\multirow{-3}{*}{\textbf{BLE}}}     & \textbf{Artificial}  & 2.16                                                & 0.80                          & \cellcolor[HTML]{C0C0C0}97.03                       \\ \hline
\end{tabular}
\label{tab:conftech}
\end{table}
Next, we analyze the confusion matrix per technology in Table \ref{tab:conftech}. It shows the percentage of predicting either no CTI, LR-WPAN or BLE in case of each actual class, obtained for all SNR and SIR using the filtered results. In general it can be seen that from the OTA, to the cable and artificial test, the model went from low to high false positive rate (predicting LR-WPAN or BLE when there was no CTI), and from low to high true positive rate. Also, the presence of LR-WPAN is more often classified as ``no CTI" than BLE, which can be explained by the fact that the power of BLE is concentrated on a smaller bandwidth, meaning that at the same SIR, BLE has a higher power spectrum density than LR-WPAN, hence its distortion on the CSI is clearer.

To assess the accuracy of the spectral location, we divided the spectrum into four 52-tone RUs, each overlapping with two BLE channels and one LR-WPAN channel. When interference is detected in the OTA test, the correct overlapping RU is detected in 99.8\% of the cases.

%Following, we determine the confusion matrix for the spectral occupancy of the interference. We do this by dividing the spectrum up into four 52-tone RUs when using OFDMA as indicated by the dotted lines in Fig. \ref{fig:BLE_LRWPAN}. The rationale behind this is that if the interference is correctly detected within those RUs, intelligent multi-user OFDMA scheduling can be applied efficiently. Note that each RU is overlapped with one LR-WPAN channel and two BLE channels. Table \ref{tab:confRU} shows that for all tests, if interference is detected, the RU index at which the interference appears is almost always correctly determined. Especially in the OTA test that has a low false positive rate, this allows to establish reliable interference countermeasures in real-life network operations.
\section{CTI-aware OFDMA Scheduling Results}
In order to show the benefit of CTI-aware OFDMA scheduling, we implement a scheduler on an openwifi AP and let the ESP32-C6 perform CTI detection. 
%In this section we will explain the implementation and the measurement setup, and then discuss the experimental results.
Openwifi \cite{jiao2020openwifi} is an open-source implementation of the IEEE 802.11 standard running on a System-on-Chip. %, compatible with Linux’ mac80211 subsystem. 
The baseband processing and low-MAC is realized on the Field Programmable Gate Array (FPGA); the driver and high-MAC run as Linux kernel modules on the on-chip ARM processor. The baseband processing for downlink OFDMA support following the IEEE 802.11ax standard has been implemented in our previous work \cite{Aslam24}. In this work, we make the necessary addition to the driver to control the RU allocation and corresponding parameters like MCS.
%The driver uses mac80211’s intermediate software queues per STA, which the scheduler can query to form an MU packet.  %For MU packets, instead of an immediate acknowledgment, separate Block Acknowledgments Requests are sent to each STA.
The STAs perform the CTI classification and send the result to the AP using Wi-Fi frames. Since the two-user OFDMA frames consisting of 2x106-tone RUs have a different CSI dimension than the 242-tone RU used previously, the model has been retrained to incorporate the new CSI dimension. % using 2x106-tone RUs as well.

For this specific OFDMA scheduler we are only interested in whether CTI is detected and its spectral location. Therefore, the AP holds a historical record of 64 interference detection results per RU and per STA within the driver. First, the AP decides upon an RU allocation based on the number of users for which it has data available. Then, users will be assigned to RUs where they experienced the least interference during the record. 
% The scheduler attempts to assign users where they experience the least amount of interference during the record, if multiple users end up in the same RU, then the scheduler iterate through the users so that they are assigned to one of the remaining relatively clean RUs.
%If this leads to conflicts, an RU will be assigned to the user who experiences the least interference in that RU. 
%Note that this method thus keeps using a certain RU allocation even when some interference-free packets are received, while still maintaining a quick response time as the classification is done using a single snapshot.
By using a single CSI snapshot combined with a historical record, the CTI detection time is relatively low, hence the scheduler can capture transient interference and quickly adapt to it.

\begin{figure}
    \centering
\includegraphics[width=\linewidth]{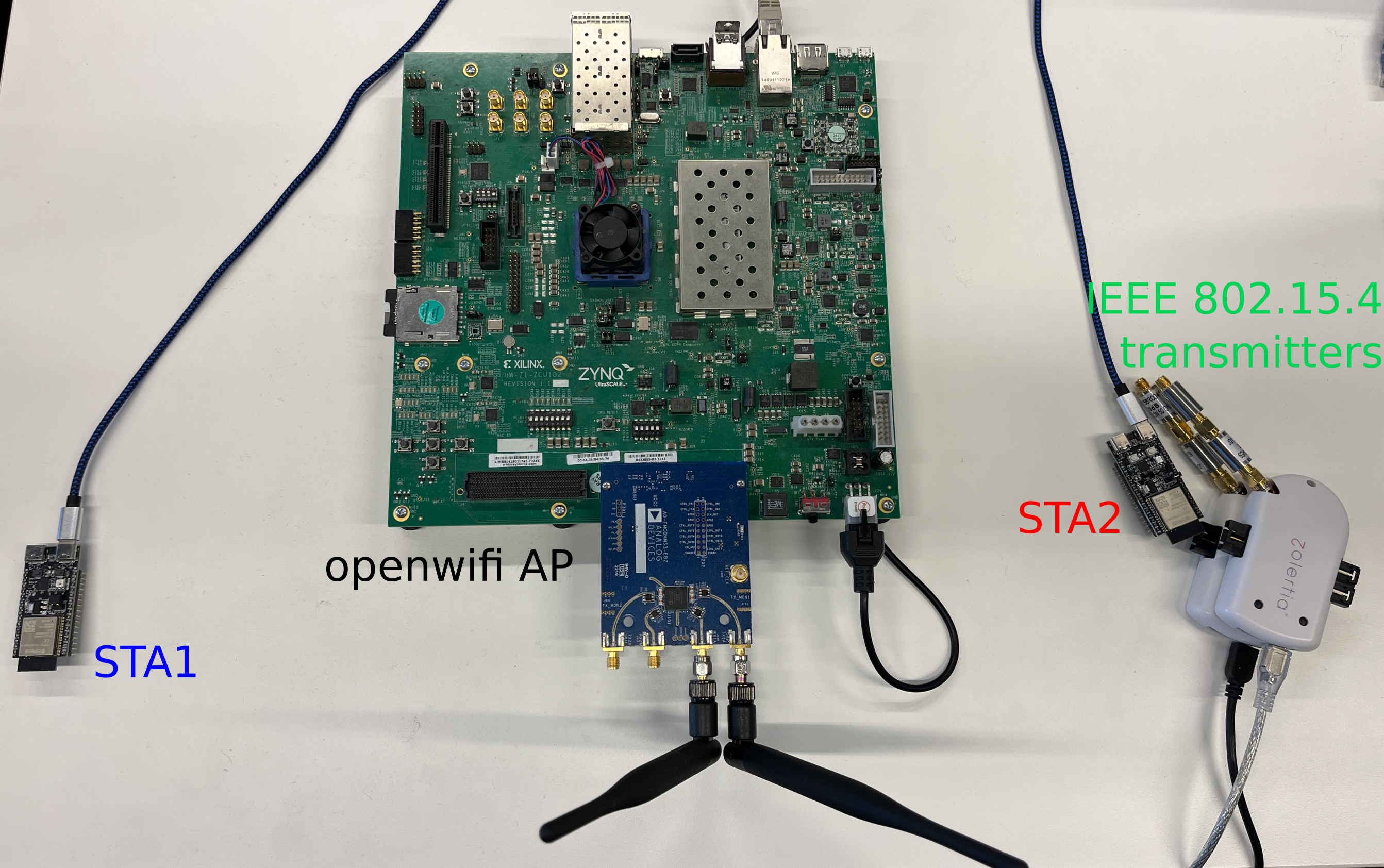}
    \caption{Experimental setup to validate the CTI-aware OFDMA scheduler on the openwifi AP, when sending downlink packets to STA1 and STA2; only STA2 is interfered by two IEEE 802.15.4 transmitters on channel 14.}
    \label{fig:CTI_setup}
\end{figure}

To validate the CTI-aware OFDMA scheduler, the experimental setup as shown in Figure \ref{fig:CTI_setup} is used. 
Namely, STA1 and STA2 (ESP32-C6 boards) are located on opposite sides of the openwifi AP, which runs on a Xilinx Zynq UltraScale+ MPSoC ZCU102 with an Analog Devices FMCOMMS3 RF front-end. Next to STA2, the two IEEE 802.15.4 transmitters\footnote{The IEEE 802.15.4 radios are the Zolertia RE-Mote revison A available in the w-iLab.t testbed: \url{https://doc.ilabt.imec.be/ilabt/wilab/hardware.html}.} are placed on top of each other. 
The openwifi AP has only limited output power, thus the STAs need to be close to the AP to stay connected. In order to create the topology where only STA2 experiences interference, the RF outputs of the LR-WPAN transmitters are connected to two \SI{20}{dB} attenuators and their output power is set to \SI{-24}{dBm}. The LR-WPAN devices continuously send request and reply packets of 80 bytes to each other, which equals \SI{3.83}{ms} of airtime per packet. Meaning that if each request is received correctly and answered with an equally long reply packet, around 51\% duty cycle is achieved. Due to packet losses, in reality the duty cycle of the interference is only around 35\%.

The Wi-Fi network runs on Wi-Fi channel 1, and the IEEE 802.15.4 transmitters use channel 14, thus overlapping with 106-tone RU 2 as shown in Figure \ref{fig:BLE_LRWPAN}. %RU 4 in case of 52-tone RUs. In the two-user scenario considered here, the MU scheduler assigns an 106-tone RU to each user, which roughly occupies either RU 1 and 2, or RU 3 and 4 in Figure \ref{fig:BLE_LRWPAN}.
%We compare the CTI-aware scheduler to a scheduler that does the opposite, meaning it will always allocate a user to the RU where it experiences the most interference, which may happen when a naive scheduler uses a fixed RU allocation. 
We compare the CTI-aware scheduler to a naive MU scheduler using a fixed RU allocation, which happens to always allocate a user to the RU where it experiences the worst interference. 
%This is referred to as the naive MU scheduler. 
Furthermore, we evaluate the performance when using SU packets. This packet format always occupies the full bandwidth, which relies on the standard Carrier Sense Multiple Access with Collison Avoidance (CSMA/CA) for CTI mitigation. 
iPerf2 is used to start UDP traffic to the two STAs simultaneously, with the default payload length of 1470 bytes. The openwifi AP is set to use a fixed MCS 7. The high MCS is chosen to clearly show the impact of CTI. For each scheduler, five tests of one minute are executed and the average throughput reported by the STAs is recorded. In addition, the measurements are repeated also when the CTI is disabled, to obtain a performance baseline.

\begin{figure}
	\centering
	\includegraphics[width=\linewidth]{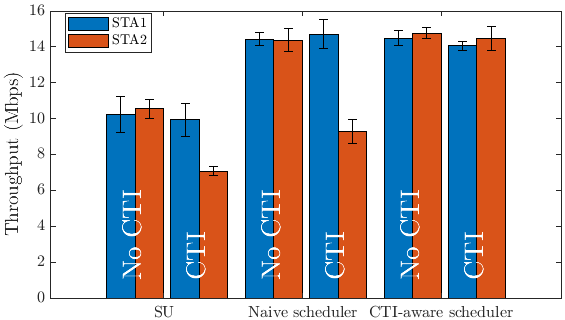}
	%\caption{Experimental throughput of different OFDMA schedulers with CTI feedback from STAs where only one STA is close to the interferer.}
    \caption{Downlink throughput of two STAs using SU packets only, MU packets using a naive RU allocation, and MU packets by CTI-aware RU allocation.}
	\label{fig:realloc_throughput}
\end{figure}

Figure \ref{fig:realloc_throughput} displays the average throughput for each STA with error bars showing the standard deviation.
In general, using SU packets results in lower total throughput, which can be partially explained by the fact that it has more overhead due to the preamble and contention time.
%\footnote{Using a 1470-byte payload per user at MCS 7 for SU, there are 11 OFDM symbols needed, while for a two-user MU packet, there are 24 OFDM symbols needed. The preamble time of an SU packet is \SI{40}{\us}, and for MU this is \SI{48}{\us}. The DIFS of \SI{34}{\us}, plus, e.g., on average 10 slot times of \SI{9}{\us} for back-off, results in \SI{124}{\us} idle time between packets. With a symbol time of \SI{16}{\us}, to send 1470 bytes to each user, it would take $2\cdot(40 + 16 \times 11 + 124) = \SI{680}{\us}$ for SU, and $48 + 16 \times 24 + 124 = \SI{556}{\us}$ for MU, which is $22\%$ more airtime for SU.}. 
For SU packets, it can be seen that when CTI is applied, the throughput towards STA1 (the blue bars) remains roughly stable, while for the STA2, it decreases by about 33\% (indicated by the red bar that drops from \SI{10.5}{Mbps} to \SI{7}{Mbps}). Roughly the same observation for the naive MU scheduler can be made, where the throughput of STA1 stays above \SI{14}{Mbps} but STA2's throughput dropped from \SI{14}{Mbps} to \SI{9}{Mbps} ($\approx$35\% drop). On the other hand, the throughput reduction for STA2 is negligible when using the CTI-aware scheduler, because the data is now sent on a portion of the spectrum that does not overlap with the CTI.

\section{Conclusions}
%In order to detect CTI on commodity IoT hardware using regular Wi-Fi traffic, 
To combat CTI, this work shows that %CTI detection and classification can be achieved using existing hardware in the radio chipsets.
one snapshot of Wi-Fi 6 CSI can provide sufficient detection accuracy for on-device machine learning-based classification of LR-WPAN and BLE interference, including their spectral location. We have shown that the CTI classification accuracy for different SNR and SIR can go up to 99\% in a scenario with high SNR and low SIR, but it drops significantly below \SI{14}{dB} SNR due to the decrease of CSI quality. 
%Furthermore, there is a threshold of SIR for which CTI can be reliably detected. 
%However, even when the technology type is sometimes confused, the detection of the spectral location of the CTI remains highly accurate. 
The overall classification accuracy is comparable or better than prior arts that use multiple CSI snapshots.
%Furthermore, although our proposed model is a general model trained on artificial data, we have found that nonidealities like local oscillator leakage and wrong carrier or sampling frequency offset estimation can hurt the performance.% or even apply online learning to optimize for a certain environment.
Furthermore, due to the artificially generated training data, we observe that non-idealities such as local oscillator leakage can mislead the model, which should be addressed in future work.

To show that CSI-based CTI classification can be effectively applied for CTI mitigation, we implement a CTI-aware OFDMA scheduler on a software-defined radio. By collecting real-time CTI detection results on the STAs, the AP correctly assigns a clean RU in the frequency domain to a user that originally experiences narrowband CTI. In this way, we prove experimentally that CTI-aware OFDMA scheduling can fully mitigate the 35\% throughput loss caused by CTI.
% For future work we suggest to not only classify the interference, but apply a regression to estimate the interference's strength, which helps to determine the mitigation technique. Furthermore, we plan to implement real-time RU puncturing to avoid ongoing interference packets when detected at the AP, which is crucial to avoid frequency-hopping CTI. 
%We intend to further reduce the footprint of the detection model and its inference time in order to optimize the spectrum access in a real-time communication link with CTI present. 
\section*{Acknowledgment}
\noindent
This work is funded by EU Horizon projects under Grant Agreement No. 101095738 (6G-SHINE) and No. 101139176 (6G-MUSICAL), and FWO SBO S003921N VERI-END.com.
\bibliographystyle{IEEEtran}

\bibliography{IEEEabrv,references.bib}

\end{document}